\documentclass[aps,pra,longbibliography,10pt,twocolumn,superscriptaddress,notitlepage,floatfix,amssymb]{revtex4-1}
\usepackage[letterpaper,hmargin={1.5cm,1.5cm},vmargin={1.5cm,2.5cm}]{geometry}
\usepackage{mathtools}
\usepackage{float}
\usepackage{amsfonts}
\usepackage{enumitem}
\usepackage{graphicx}
\usepackage{color}
\newcommand*{\ham}{\hat{H}}

\newcommand*{\cre}[2][a]{\hat{#1}_{#2}^{\dagger}}	

\newcommand*{\ann}[2][a]{\hat{#1}_{#2}}

\newcounter{tableeqn}[table]

\newcounter{tablesubeqn}[tableeqn]

\begin{document}
\title{Observation of Three-Photon Spontaneous Parametric Downconversion in a Superconducting Parametric Cavity}

\author{C.W. Sandbo Chang}
\affiliation{Institute for Quantum Computing and Electrical and Computer Engineering, University of Waterloo, Waterloo, Canada}
\author{Carlos Sab{\'\i}n}
\affiliation{Instituto de F{\'i}sica Fundamental, CSIC, Serrano, 113-bis, 28006 Madrid, Spain}
\author{P. Forn-D\'{i}az}
\affiliation{Institut de F\'isica d'Altes Energies (IFAE), The Barcelona Institute of Science and Technology
(BIST), Bellaterra (Barcelona) 08193, Spain}
\affiliation{Barcelona Supercomputing Center - CNS, Barcelona 08034, Spain}
\author{Fernando Quijandr\'{\i}a}
\affiliation{Microtechnology and Nanoscience, MC2, Chalmers University of Technology, SE-412 96 G\"oteborg, Sweden} 
\author{A.M. Vadiraj}
\affiliation{Institute for Quantum Computing and Electrical and Computer Engineering, University of Waterloo, Waterloo, Canada}
\author{I. Nsanzineza}
\affiliation{Institute for Quantum Computing and Electrical and Computer Engineering, University of Waterloo, Waterloo, Canada}
\author{G. Johansson}
\affiliation{Microtechnology and Nanoscience, MC2, Chalmers University of Technology, SE-412 96 G\"oteborg, Sweden}
\author{C.M. Wilson}
\affiliation{Institute for Quantum Computing and Electrical and Computer Engineering, University of Waterloo, Waterloo, Canada}

\email{chris.wilson@uwaterloo.ca}

\begin{abstract}
Spontaneous parametric downconversion (SPDC) has been a key enabling technology in exploring quantum phenomena and their applications for decades. For instance, traditional SPDC, which splits a high energy pump photon into two lower energy photons, is a common way to produce entangled photon pairs. Since the early realizations of SPDC, researchers have thought to generalize it to higher order, \textit{e.g.}, to produce entangled photon triplets. However, directly generating photon triplets through a single SPDC process has remained elusive. Here, using a flux-pumped superconducting parametric cavity, we demonstrate direct three-photon SPDC, with photon triplets generated in a single cavity mode or split between multiple modes. With strong pumping, the states can be quite bright, with flux densities exceeding 60 photon/s/Hz. The observed states are strongly non-Gaussian, which has important implications for potential applications. In the single-mode case, we observe a triangular star-shaped distribution of quadrature voltages, indicative of the long-predicted ``star state".  The observed star state shows strong third-order correlations, as expected for a state generated by a cubic Hamiltonian. By pumping at the sum frequency of multiple modes, we observe strong three-body correlations between multiple modes, strikingly, in the absence of second-order correlations. We further analyze the third-order correlations under mode transformations by the symplectic symmetry group, showing that the observed transformation properties serve to ``fingerprint" the specific cubic Hamiltonian that generates them. The observed non-Gaussian, third-order correlations represent an important step forward in quantum optics and may have a strong impact on quantum communication with microwave fields as well as continuous-variable quantum computation.  
\end{abstract}

\date{\today}

 \pacs{42.50.Gy, 85.25.Cp, 03.67.Hk}

\maketitle

\section{Introduction}
For over thirty years, spontaneous parametric downconversion (SPDC) has been a workhorse for quantum optics. Famously known as a process which generates photons in pairs from a single pump, it has had a crucial role in fundamental tests of quantum theory \cite{Friberg1985,Tittel1998,Burnham1970} as well as many applications in quantum information processing \cite{Tittel2000,Walther2005,Didier2015}. Spanning frequencies from optical to microwave, SPDC has a central role, for instance, in quantum-limited amplifiers \cite{Yamamoto2008} and sources of nonclassical light, including squeezed states, Fock states \cite{Cooper2013}, and entangled photon pairs \cite{Kwiat1995}. This broad and important set of phenomena has been referred to as ``two-photon quantum optics"\cite{Caves1985,Schumaker1985}. From early days, generalizations of the standard two-photon SPDC have been explored, but this endeavor has proven difficult both theoretically \cite{Fisher1984, Hillery1984,Braunstein1987, Braunstein1990,Hillery1990,Banaszek1997,Felbinger1998,Olsen2002,Marshall2015,Arzani2017} and experimentally \cite{Bencheikh2007,Gonzalez2018}. 

Even the pursuit of the next order, three-photon SPDC which would create photon triplets, has gone on for decades \cite{Borshchevskaya2015,Corona2011,Akbari2016,Cavanna2016}. This process has been studied theoretically in the context of \textit{generalized} squeezing \cite{Braunstein1987,Zelaya2018,Fisher1984}, where the quantum vacuum is shaped by the action of nonquadratic Hamiltonians \cite{Ding2018}. The non-Gaussian nature of these higher-order squeezed states could make them a resource for universal quantum computation with linear optics \cite{Ghose2005,Gu2009,Takagi2018}. In particular, cubic squeezing is one path to generate the ``magic" cubic-phase state of Gottesman, Kitaev, and Preskill \cite{Gottesman2001} that enables error-correction in continuous-variable (CV) quantum computing. Three-photon SPDC has also been studied as a source of more sophisticated, three-photon entangled states, such as GHZ states \cite{Agne2017}, as well as heralded entangled pairs \cite{Hamel2014}.  These states would be useful, e.g., for novel quantum communication protocols such as quantum secret sharing \cite{Hillery1999}. Despite the great potential of multiphoton downconversion and generalized squeezing, their experimental demonstration has remained elusive. 

Here, we report an experimental implementation of three-photon SPDC producing generalized squeezed states, in particular, trisqueezed states.  This is done using a flux-pumped, superconducting parametric resonator. By the choice of pump frequency, we can alternately produce degenerate three-photon downconversion to a single mode or nondegenerate three-photon downconversion to three distinct modes. Further, we can produce a hybrid version to two modes, where two photons are degenerate and one is nondegenerate.  Our triplet source is bright, producing a propagating photon flux with a flux density controllable from less than 1 to greater than 60 photon/s/Hz over a bandwidth of hundreds of kHz, well surpassing any records to date for photon triplets and comparable to ordinary two-photon downconversion (TPDC) experiments \cite{Flurin2015,SandboChang2018}. The high flux allowed us to perform detailed analysis of the novel phase-space distributions and strongly non-Gaussian statistics of the states. For instance, we clearly see strong three-body correlations in the absence of normal two-body correlations (covariance). The symmetry properties of these correlations allow us to ``fingerprint" the Hamiltonians that created them, clearly demonstrating that states are generated by a family of pure cubic Hamiltonians with little contamination from typical quadratic processes.  These results form the basis of an exciting new paradigm of three-photon quantum optics. 

\begin{figure}[!htb]
\center
\includegraphics[width=1\linewidth]{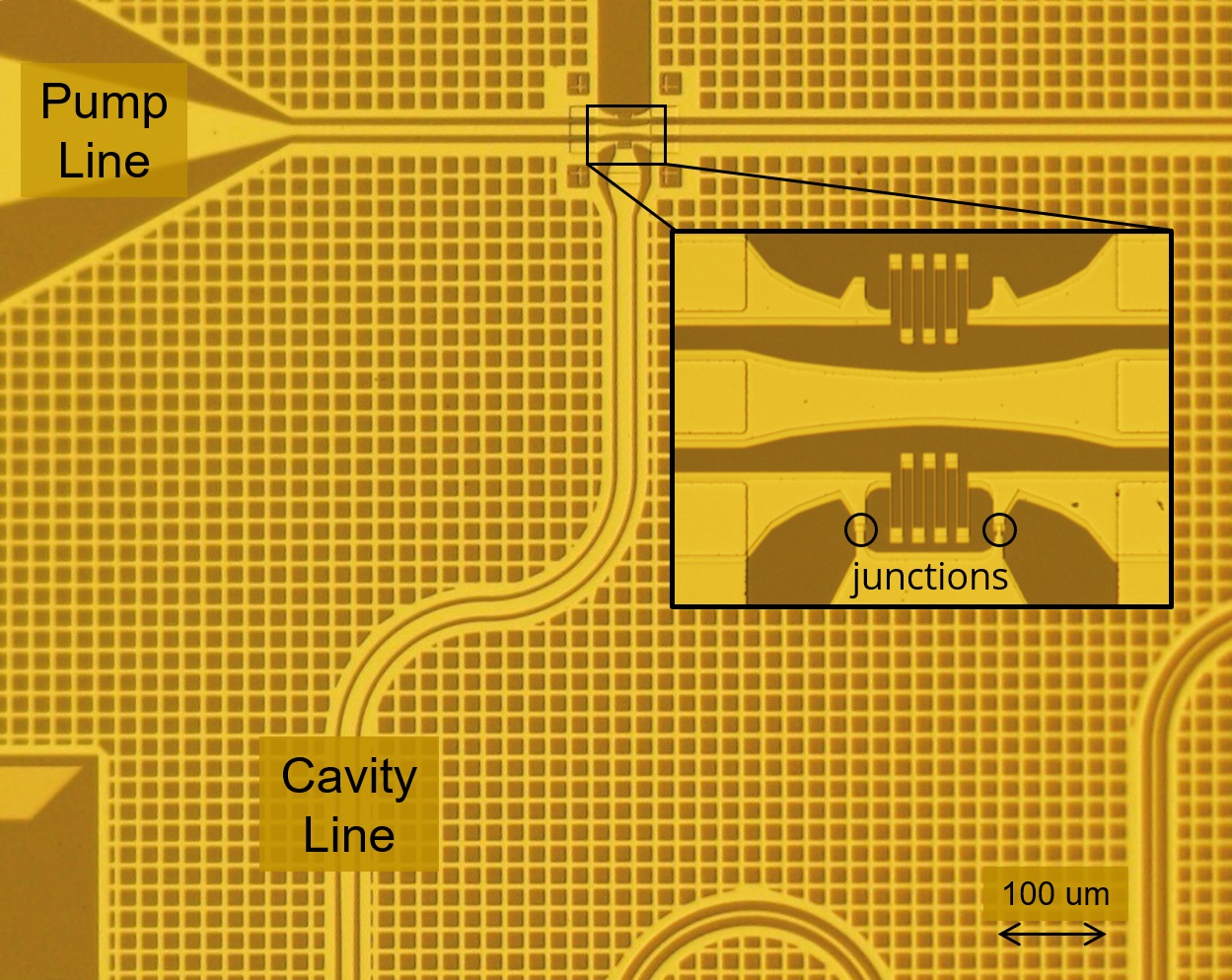}
\caption{Meandered SQUID design with improved pump coupling. The narrow, meandered path is shared between the SQUID and the ground plane of the pump line. The kinetic inductance of the meander then directly couples the pump current in the ground plane to the SQUID phase. The center conductor of the pump line is narrowed to \textit{reduce} its magnetic coupling to the SQUID, which contributes with the opposite sign of the kinetic coupling. Following microwave simulations, a mirrored structure is patterned in the top ground plane, such that the ground current divides evenly between the two ground planes. Compared to previous designs, we found a reduction in the pump strength required for TPDC by three orders of magnitude.  The reduction greatly reduces the spurious effects of strong pumping and was critical to realizing three-photon downconversion.}
\label{Device}
\end{figure}

\section{Background}
Since the first applications of SPDC were realized with a quadratic Hamiltonian, generalizing squeezing with higher-order Hamiltonians has been a topic of intense research \cite{Fisher1984, Braunstein1987, Hillery1990} .  In the case of single-mode SPDC, a generalized squeezing process can be described by a SPDC Hamiltonian of order $k$ by
\begin{align*}
\ann[H]{}=\hbar g_k \left(\alpha^*\ann[a]{}^k+\alpha\ann[a]{}^{\dagger k}\right),
\end{align*}
where $\alpha$ represents the pump (under the parametric approximation), $g_k$ is the $k$-th order coupling constant, and $\ann{}$ is the annihilation operator of the single mode. First discussed in the literature in the mid-1980s, early work \cite{Fisher1984} concluded that generalized squeezing processes were unphysical. Subsequent work \cite{Braunstein1987} showed that the issues were mainly mathematical and calculated the first phase-space distributions of generalized states using novel analytical continuation techniques. It was also later shown that the apparent divergences that appeared under the parametric approximation were removed when the pump was quantized \cite{Hillery1990}.  Despite the early theoretical challenges, continued research was motivated by the novel non-Gaussian statistics and nonclassical properties, including Wigner negativity \cite{Banaszek1997}, absent in conventional squeezed states but predicted in the higher-order squeezed states. These many-photon states continue to be attractive and potentially useful in various quantum information processing applications. 

Meanwhile, despite decades of studies of the theoretical aspects of generalized squeezing, the experimental implementation of an elementary SPDC process capable of generating three or more photons has remained an outstanding challenge, largely attributable to the relatively weak nonlinearities in optical materials \cite{Douady2004}. While experiments such as cascaded SPDC and photon generation by a quantum dot have successfully generated three-photon correlations \cite{Hamel2014,Khoshnegar2017,Agne2017a}, the triplet generation rate in these schemes was small, making detailed analysis of the three-photon states or the underlying Hamiltonian difficult. These schemes also lack explicit control over the down-converted frequencies, which made it difficult, for instance, to generate single-mode photon triplets. More recently, classical period-tripling parametric oscillations have been demonstrated with a parametric superconducting cavity, with a third-subharmonic oscillation observed when a strong flux pump was applied \cite{Svensson2017}. There have also been recent theoretical studies of quantum effects in period-tripled parametric oscillators \cite{Lorch2019}.

Superconducting parametric cavities and resonators, based on SQUIDs, are an important device family widely used for applications requiring TPDC \cite{Eichler2011,Flurin2012,Flurin2015}. In these devices, leading-order terms of the full cosine nonlinearity of the SQUID are used to generate two-photon downconversion. As described below in more detail, for this work, we modified and extended these devices to access higher-order nonlinearities of the SQUID. 

Our device is a quarter-wavelength coplanar waveguide resonator terminated by a SQUID at one end, with the other end capacitively overcoupled $(Q\approx7000)$ to a nominally $Z_0=50\Omega$ line. The fundamental mode has a relatively low frequency of around 1 GHz such that there are three higher-order modes accessible within our 4-8 GHz measurement bandwidth. Impedance engineering \cite{Zakka-Bajjani2011, Simoen2015, SandboChang2018} is used to make the mode spacing nondegenerate, yielding mode frequencies of approximately $4$, $6$ and $7$ GHz. Parametric processes are driven by a microwave pump inductively coupled to the SQUID, modulating the boundary condition of the cavity. By exploiting kinetic inductance (Fig.~\ref{Device}) and other improvements, we increased the pump coupling by roughly 30 dB compared to previous designs \cite{Simoen2015}, a crucial element in realizing higher-order parametric processes.

Many of the parametric cavities and resonators described above employed a nominally symmetric SQUID, where the two component Josephson junctions are designed to have the same Josephson energy. This comes with various advantages such as maximizing the frequency tunability of the cavities. However, it also suppresses cubic interactions between the cavity modes due to the even symmetry of the pure cosine nonlinearity of the symmetric SQUID. To access cubic (odd) nonlinearities we must therefore break the symmetry of the SQUID.  To understand this in more detail, we can write the general sinusoidal Hamiltonain of an asymmetric SQUID, coupling the cavity flux, $\hat{\Phi}_{cav}$ to the external pump flux $\hat{\Phi}_{ext}$ as  
\begin{align}
\hat{H}_{\text{SQ}}&=E_J(\hat{\Phi}_{ext})\cos(2\pi\hat{\Phi}_{cav}/\Phi_0-\alpha),\\
E_J(\hat{\Phi}_{ext})&=\sqrt{E_{J,1}^2+E_{J,2}^2+2E_{J,1}^2E_{J,2}^2\cos\left(2\pi\frac{\hat{\Phi}_{ext}}{\Phi_0}\right)},\label{EqSinusoid}\\
\alpha&=\arctan\left[\tan\left(\pi\frac{\hat{\Phi}_{ext}}{\Phi_0}\right)\frac{E_{J,1}-E_{J,2}}{E_{J,1}+E_{J,2}}\right].
\end{align} 
$E_J(\ann[\Phi]{ext})$ is the flux-dependent Josephson energy of the SQUID.  $E_J$ is tuned by the external flux $\hat{\Phi}_{ext}=\hat{\Phi}_{p}+\Phi_{bias}$, composed of the AC pump flux and DC flux bias. $E_{J,i}$ are the Josephson energies of the individual junctions.
$\alpha$ is the effective cavity flux bias arising from the SQUID asymmetry, which is necessary to access the cubic nonlinearity of the SQUID.
For the experiments presented here, we chose the ratio of the SQUID junction areas to be 1:1.7.  

We would now like to understand how the external pump affects the cavity modes. We start by expanding $E_J(\hat{\Phi}_{ext})$ to leading order in $\hat{\Phi}_{p}$ around the working point $\Phi_{bias}$.  Assuming the pump to be a large-amplitude coherent state, we further apply the parametric approximation to the pump, representing it by the classical amplitude $\beta_{p}$.  By now expanding the sinusoidal potential Eq.~\eqref{EqSinusoid} in $\hat{\Phi}_{cav}$ we arrive at the interaction Hamiltonian
\begin{align}
\ham_{\text{int}} = \sum_k \ham_{\text{k}}= \beta_{p} \sum_k \hbar g_k\left(\ann[a]{1}+\cre[a]{1}+\ann[a]{2}+\cre[a]{2}+\ann[a]{3}+\cre[a]{3}\right)^k,	\label{Hint}
\end{align} 
where the usual bosonic operators $\ann{i},\cre{i}$ correspond to the three cavity modes considered, with resonant frequencies  $f_i$. The familiar quadratic interactions, which produce TPDC, arise from keeping just the $k=2$ term of this expression \cite{Johansson2010}, which is the lowest nonvanishing order if the SQUID is symmetric. 

This Hamiltonian obviously contains a wide variety of quadratic, cubic and higher-order terms. However, the presence of the pump gives us great flexibility in selectively enhancing the strength of certain desired terms. In the interaction picture, the various terms of $\ham_{\text{int}}$ have distinct time dependencies, oscillating at frequencies such as $3f_1$ for the term $\ann[a]{1}^3$. With the appropriate choice of pump frequency, $f_p$, and the application of the rotating-wave approximation, we quickly see that we can selectively activate a desired set of parametric processes between the cavity modes, including cubic interactions.  Importantly, as is standard in the parametric approximation, the terms with time dependencies that match the pump frequency have their strength enhanced by the pump amplitude, that is, they have an effective interaction strength $|g| = |\beta_p|g_k$.  Together with cavity modes that can be easily designed, this device gives us a rich toolbox of both conventional and novel parametric processes, which can be explored with a high degree of flexibility. In this work, we will limit our discussion to cubic Hamiltonians that produce three-photon SPDC, generating photons in one, two and three modes.  These Hamiltonians and their corresponding pump frequencies are shown in Table~\ref{effIntHamTable}.  We note that we have observed other Hamiltonians, such as three-photon generalizations of beamsplitter interactions \cite{Leghtas2015}, but we will not explore them here.


\begin{table*}[htb!]
\begin{tabular}{c|l|cccc|ll}
SPDC & \multicolumn{1}{c|}{Combinations} & \multicolumn{4}{c|}{Frequency [GHz]} & \multicolumn{2}{c}{Effective Hamiltonians} \\ \hline
 & \multicolumn{1}{c|}{} & Pump & Mode 1 & Mode 2 & Mode 3 & \multicolumn{1}{c}{} &  \\
Single-mode & $f_{p1}=3\times f_1$ & 12.6 & 4.2 & - & - & $\ham_{\text{1M}}=\hbar g\left(\ann[a]{1}^3+\ann[a]{1}^{\dagger3}\right)$   \\
Two-mode & $f_{p2}=2\times f_1+f_2$ & 14.5 & 4.2 & 6.1 & - & $\ham_{\text{2M}}=\hbar g\left(\ann[a]{1}^2\ann[a]{2}+\ann[a]{1}^{\dagger2}\cre{2}\right)$   \\
Three-mode & $f_{p3}=f_1+f_2+f_3$ & 17.8 & 4.2 & 6.1 & 7.5 & $\ham_{\text{3M}}=\hbar g\left(\ann[a]{1}\ann[a]{2}\ann[a]{3}+\cre{1}\cre{2}\cre{3}\right)$  
\end{tabular}
\caption{Cubic SPDC Hamiltonians explored in this work. Each row represents a distinct process and individual experiment, where we pump the SQUID at the specified frequency, driving three-photon downconversion to one, two or three cavity modes. We choose which Hamiltonian to explore simply by setting the pump frequency. }
\label{effIntHamTable}
\end{table*}


\section{Measurements}
\subsection{Three-photon SPDC to a single mode}
By pumping at the appropriate frequency, we have experimentally realized the three cubic SPDC Hamiltonians listed in Table~\ref{effIntHamTable}. We measure the propagating output state of the cavity by first amplifying it with a cold HEMT amplifier and then a room temperature amplifier chain. The system is absolutely calibrated using a shot noise tunnel junction, as described in detail in Ref.~\cite{SandboChang2018}, giving calibrated values for the system gain, the system noise temperature, and the physical temperature of the input state.  The amplified signal is split at room temperature and the three modes are measured simultaneously using heterodyne detection, giving the field quadratures for the modes. In the rest of the paper, we denote the raw, room-temperature quadratures as $\ann[I]{}=\sqrt{G}\left(\ann{}+\cre{}\right)$ and $\ann[Q]{}=-\sqrt{G}i\left(\ann{}-\cre{}\right)$, and the calibrated quadratures, referred to the output of the cavity, as $\ann[x]{}=\ann{}+\cre{}$, $\ann[p]{}=-i\left(\ann{}-\cre{}\right)$. The scaling factor $G$ includes the calibrated system gain. To quantify the downconverted signal emitted by the cavity, we define the photon flux density, $F(\omega)$ defined by $\langle \cre{}(\omega)\ann{}(\omega') \rangle = F(\omega)\delta(\omega-\omega')$, which is the average number of photons/s/Hz propagating in the transmission line.

We start by exploring three-photon SPDC to a single cavity mode. We use the mode at approximately $f_1=4.2$ GHz, by applying a pump tone at three times the cavity mode frequency, \textit{i.e.}, $f_p=12.6$ GHz. This activates the Hamiltonian $\ham_{\text{1M}}$ in Table~\ref{effIntHamTable}. The effects of this single-mode Hamiltonian have been theoretically studied the most among the three Hamiltonians in Table~\ref{effIntHamTable}. Ref.~\cite{Banaszek1997} predicted that phase-space interference between the downconverted photon triplets would lead to a highly non-Gaussian Wigner function with the profile of a three-pointed star, with the three arms having triangular symmetry. Due to this distinct pattern, this single-mode trisqueezed state has also been called a star state \cite{Braunstein1987}. 

\begin{figure*}[htp!]
\center
\includegraphics[width=1\linewidth]{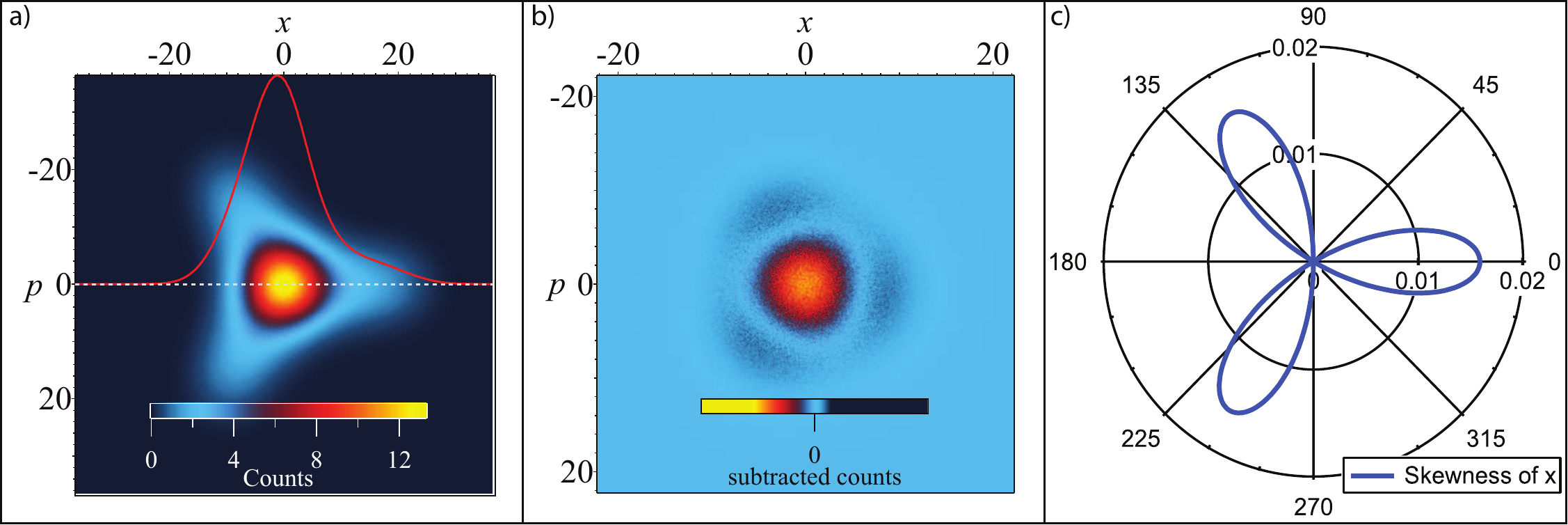}
\caption{Measured histograms of trisqueezed states in a single mode. 
a) A histogram of a bright trisqueezed state with $F\approx66$. The triangular symmetry of the histogram is apparent, clearly indicating its non-Gaussian character. The inset shows the projection of the the 2D histogram onto the x quadrature, which displays a significant amount of asymmetry (skewness), despite the system noise.  b) A histogram of a trisqueezed state with $F\approx 1$ after subtracting the system noise histogram. The triangular symmetry is still apparent despite of the low photon flux. c) A polar plot of the skewness, $\gamma_\varphi$, of the generalized quadrature $x_\varphi$ (see text) as a function of the phase angle, $\varphi$. The plot is produced from the same raw data as (b) but without subtracting the system noise.  This essentially measures the asymmetry of the quadrature distribution with respect to the symmetry axis perpendicular to $x_\varphi$. We see that the triangular symmetry of the underlying state is very visible even though it is completely obscured in a histogram of the raw data.  This suggests that $\gamma_\varphi$ is a good way to analyze the non-Gaussian character of our trisqueezed states.  This is particularly true as the skewness of the Gaussian system noise is zero (within our measurement error). 
}
\label{singleModeFig}
\end{figure*}

To look for experimental evidence of the star state, we start by simply examining phase-space histograms of the calibrated quadratures of the cavity output at $f_1$. The cavity input state is near vacuum, at a calibrated temperature of $30$ mK.  Fig.~\ref{singleModeFig} shows the results for two different pump strengths. Fig.~\ref{singleModeFig}a shows the histogram of a strongly pumped state with  $F=66$. For this bright state, the triangular symmetry of the generated star state is readily apparent, even on top of the Gaussian system noise measured at $F=35$, illustrating the strong cubic interaction achieved. In Fig.~\ref{singleModeFig}b, we also show data for a weakly pumped state with $F\approx1$.  In this case, the raw histogram is dominated by the system noise.  However, the triangular symmetry of the state can still be observed when we subtract the (unpumped) system noise histogram from the pumped signal histogram. 


Clearly, the phase-space distributions of our single-mode trisqueezed states are radically different from a two-photon Gaussian squeezed state, which we expect to lead to very different statistics. We will use these statistics to make a more quantitative comparison between the theoretical predictions and our experimental results. Generally, we expect higher-order moments beyond the second-order variance and covariance to be significant. In particular, as we expect our interaction Hamiltonians to be cubic, three-point correlators are of interest. Accordingly, we quantify the non-Gaussian character of our measured quadratures using the third standardized moment, also called the skewness, which for the random variable $y$ is given by $\gamma(y) = \frac{\overline{(y-\overline{y})^3}}{\sigma_y}$, where the overbar represents the expectation value, $\overline{y}$ is the mean, and $\sigma_y^2$ is the variance of $y$. (Below, we generally assume $\overline{y} = 0$.) Roughly, the skewness measures the asymmetry of the distribution of $y$ and is zero for Gaussian variables. We note that in what follows, we will use the notational convention that, \textit{e.g.}, $x$ represents a classical measurement record associated with the observable operator $\hat{x}$. Further, we will take quantities like $\langle x^3\rangle$ to represent classical time-averages of measurement records, while $\langle \hat{x}^3\rangle$ represents the quantum correlator. 

Looking at Fig.~\ref{singleModeFig}, it is clear that our measured quadrature distributions are asymmetric, but also that the asymmetry is not invariant under phase rotations.  That is, the skewness of $\ann[x]{}$ and $\ann[p]{}$ are not generally the same.  To study the transformation properties of the skewness, we can define the generalized quadrature $ \ann[x]{\varphi} = \ann[x]{}\cos\varphi - \ann[p]{}\sin\varphi$.  We can then study the skewness of the measurements of $\ann[x]{\varphi}$, defining for simplicity $\gamma_\varphi = \gamma(x_\varphi)$. Essentially, $\gamma_\varphi$ measures the asymmetry of the quadrature distribution with respect to the symmetry axis perpendicular to the direction of  $\ann[x]{\varphi}$.  We can also associate $\gamma_\varphi$  with the three-point quantum correlator $\left\langle\ann[x]{\varphi}^3 \right\rangle$. Fig.~\ref{singleModeFig}c shows a polar plot of $\gamma_\varphi$ as a function of $\varphi$, for the data measured with $F\approx1$.  Unlike Fig.~\ref{singleModeFig}b, we do not subtract the amplifier noise before calculating $\gamma_\varphi$, as the amplifier noise is expected to be symmetric and not contribute.  We can make a number of comments on Fig.~\ref{singleModeFig}c.  First, we see that the signal-to-noise ratio in measuring $\gamma_\varphi$ is quite good, despite it being a higher-order statistic. Second, we see that the triangular symmetry of the underlying star state is quite apparent even though, in the raw data, it is completely obscured by the amplifier noise. Next, we can conclude from the strong nodes at $\varphi = {\textstyle \frac{2\pi}{3}}(n + 1/2)$ that the skewness of the amplifier noise is indeed small. Finally, we can observe that the nodes correspond to angles where the symmetry plane aligns with a lobe of the star, while the antinodes correspond to the symmetry plane being perpendicular to a lobe. Overall, we observe that measuring $\gamma_\varphi$ appears to be a useful way to characterize the non-Gaussian character of the single-mode trisqueezed states.

\subsection{Three-photon SPDC to multiple modes}
Now, we look at multimode trisqueezed states.  As previously mentioned, this includes two-mode states and three mode states. Once again, these states can be produced by the appropriate choice of the pump frequency, naturally given by the conservation of energy (see Table~\ref{effIntHamTable}). Similar to the single-mode trisqueezed state, we expect significant third-order statistics in the output states. Thus, we statistically characterize the two-mode and three-mode trisqueezed states using the so-called coskewness of $A,B,C$ defined as $\gamma_{ABC} = \frac{\overline{ABC}}{\sigma_A\sigma_B\sigma_C}$, here assuming the measurements are mean zero. Now, we can associate this statistical measure with the three-point quantum correlators $\langle\ann[A]{}\ann[B]{}\ann[C]{}\rangle$, where $\ann[A]{},\ann[B]{}$ and $\ann[C]{}$ are some quadratures of the multimode trisqueezed states. 


\begin{figure}[t]
\center
\includegraphics[width=1\linewidth]{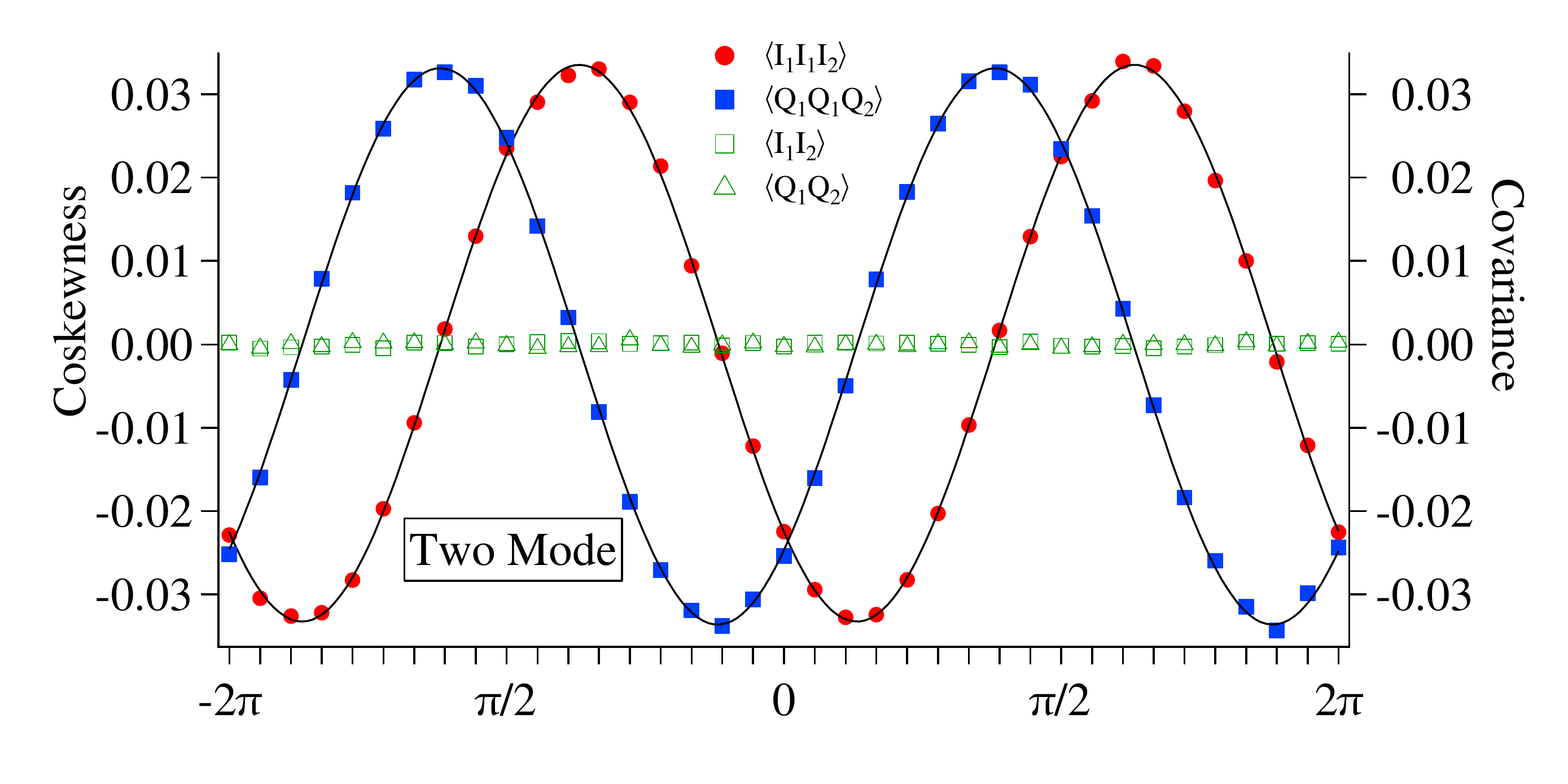}
\includegraphics[width=1\linewidth]{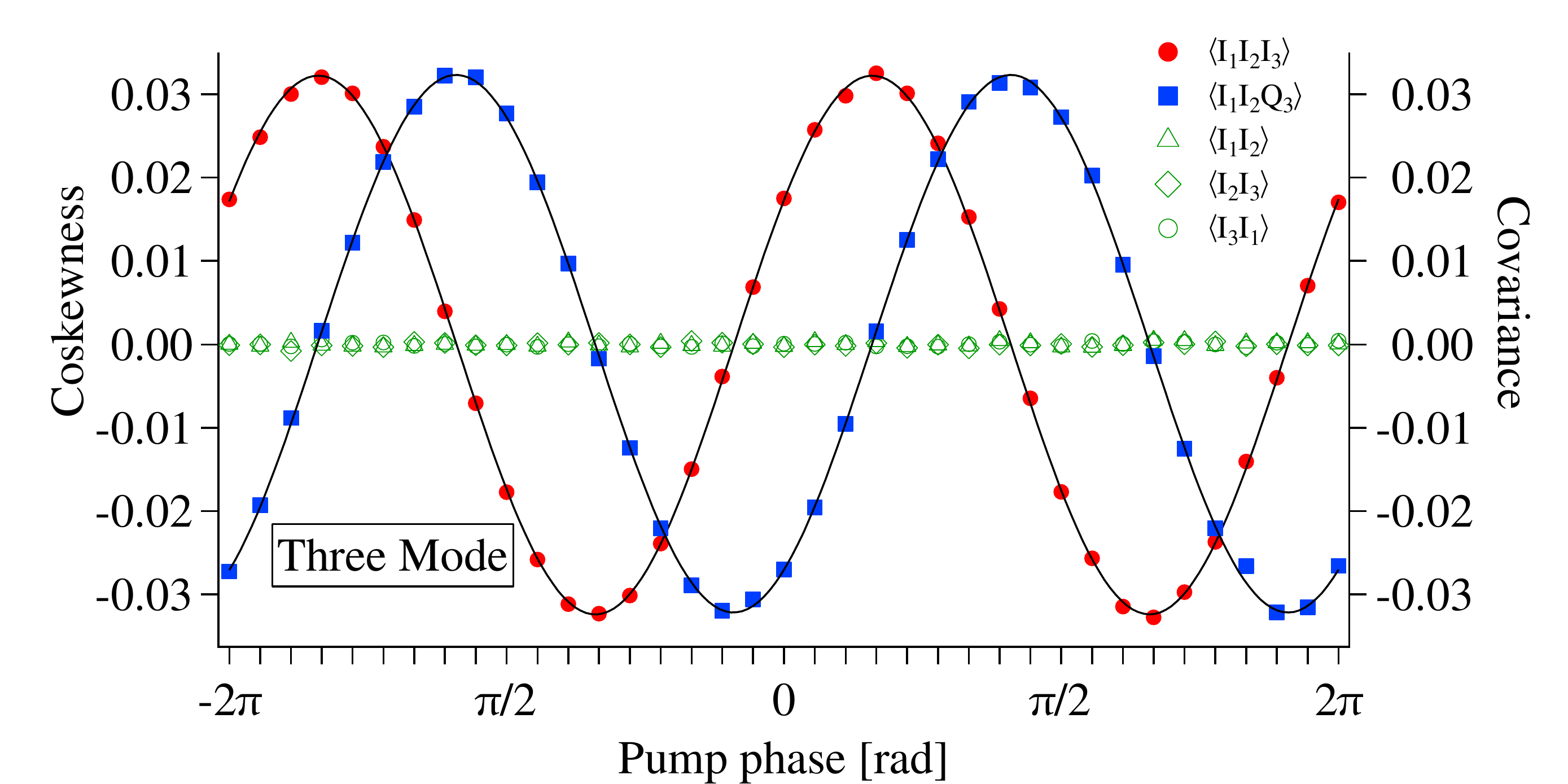}
\caption{Covariance and coskewness of three-photon SPDC to (top) two modes and (bottom) three modes measured at room temperature. In the two-mode case, we have a degeneracy between two of the three generated photons. This leads to one mode participating twice in the nonzero coskewness term. For our choice of pump phase, the only significant coskewness terms are then $\langle I_1^2 I_2 \rangle$ and $\langle Q_1^2 I_2 \rangle$. In the three-mode case, the only nonzero coskewness term must contain all three modes, \textit{e.g.}, $\langle I_1 I_2 I_3 \rangle$. To show that the observed coskewness is coherently generated by the pump, we sweep the pump phase from $-2\pi$ to $2\pi$ and observe the effect.  We see a clear oscillation in the coskewness. The oscillations are fit well by a sinusoid. Meanwhile, all covariance terms are essentially zero throughout the sweep, indicating that the generated states does not contain second-order correlations.}
\label{PumpSweepCoskewness}
\end{figure}

\begin{figure}[t]
\center
\includegraphics[width=0.9\linewidth]{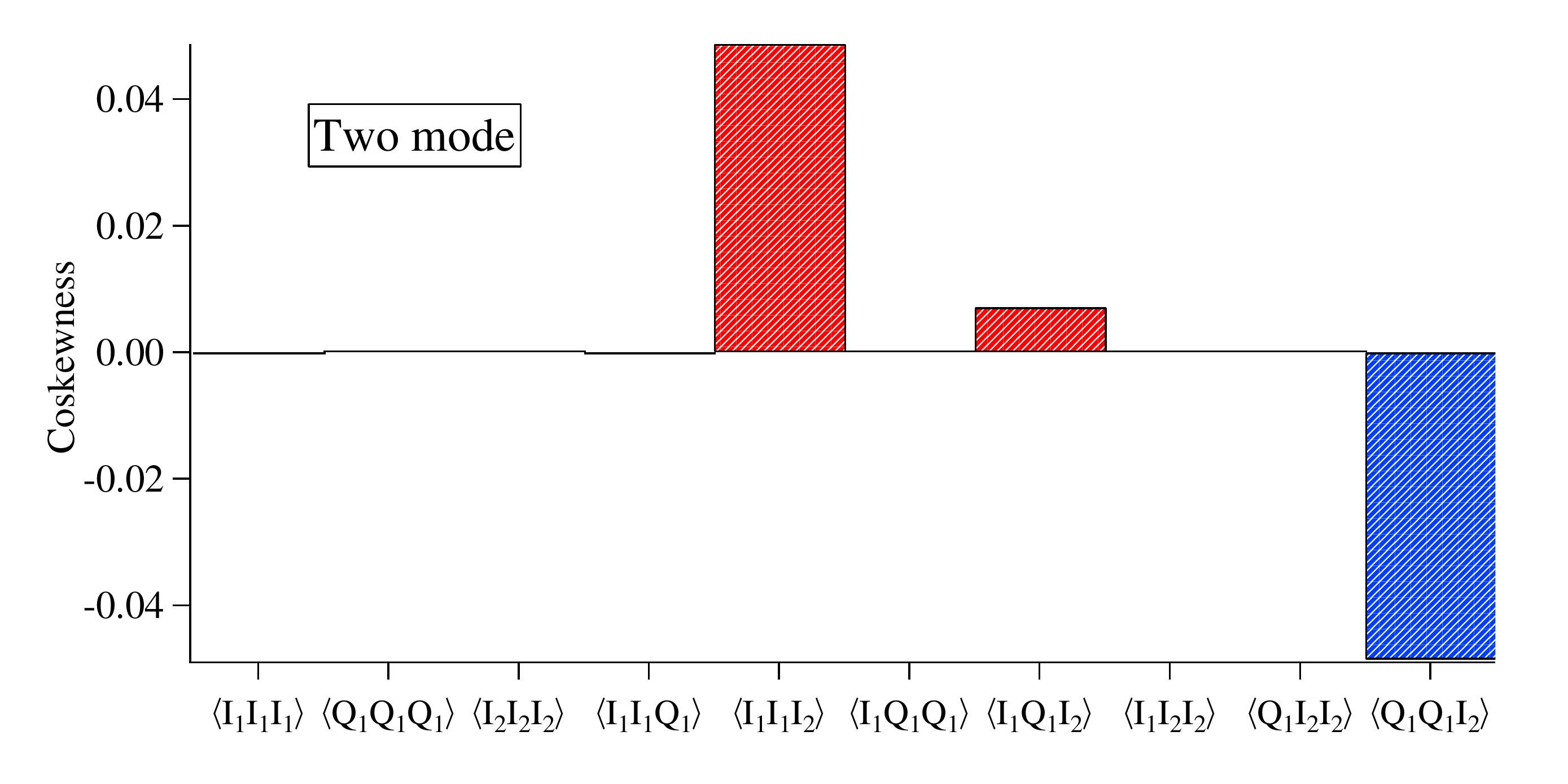}
\includegraphics[width=0.9\linewidth]{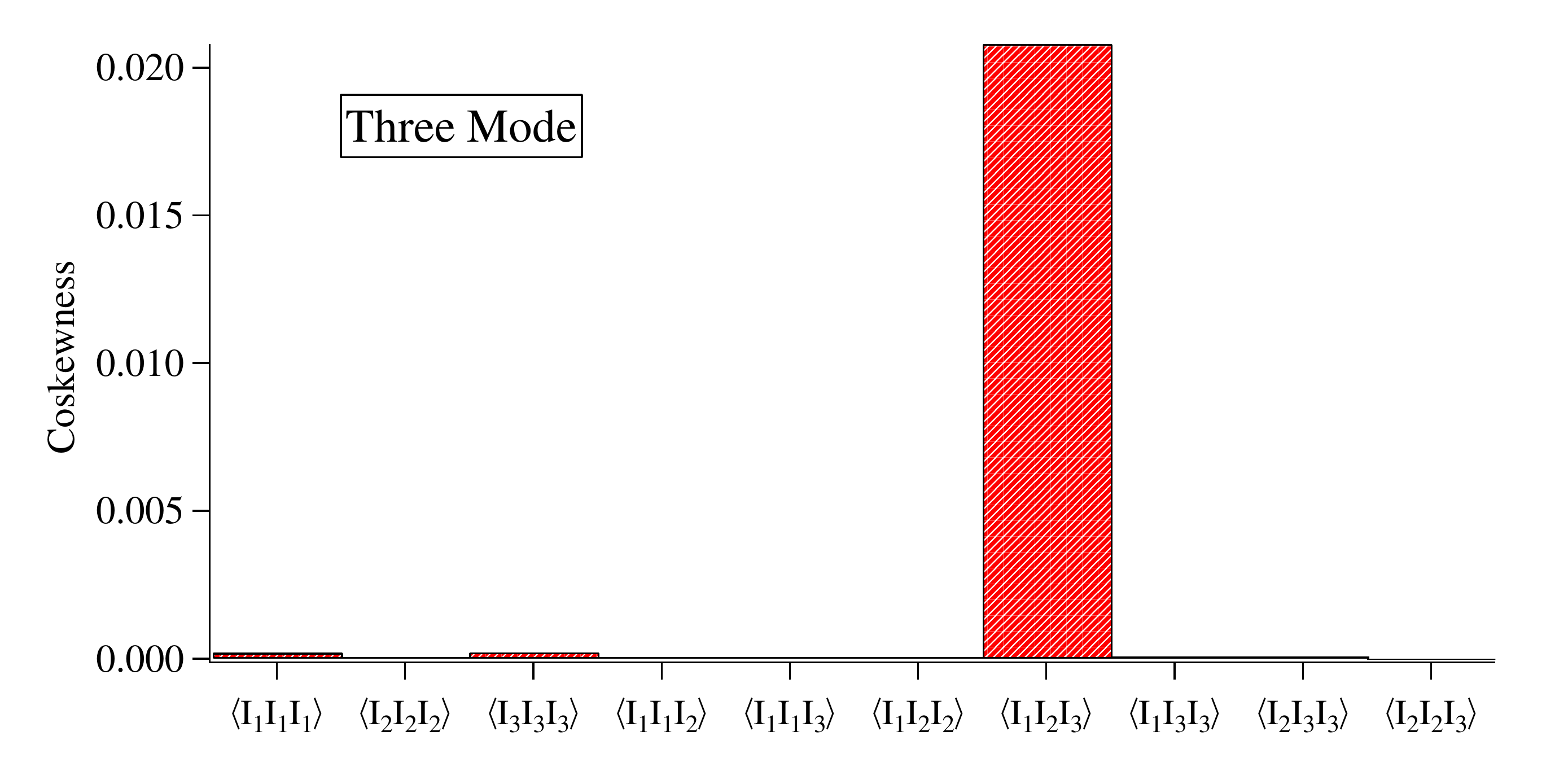}
\caption{Measured coskewness in (top) the two-mode case and (bottom) the three-mode case. In the two mode case, we considered the three quadratures $\ann[I]{1}$, $\ann[Q]{1}$ and $\ann[I]{2}$ as a representative sample.  With the chosen pump phase, $\langle I_1^2 I_2 \rangle$ and $\langle Q_1^2 I_2 \rangle$ are expected to be the only nonzero coskewness terms. This is well reflected in the data. The small amplitude at $\langle I_1 Q_1 I_2 \rangle$ is caused by a phase mismatch between the pump and the digitizers in the experiment. In the three-mode case, we consider the $\ann[I]{}$ quadratures of all three modes. As expected, the only nonzero term includes all three modes, i.e. $\langle I_1 I_2 I_3 \rangle$.}
\label{InteractionList}
\end{figure}

While pumping, the amplified output signal of the cavity was split at room temperature and the three modes measured simultaneously by three digitizers to obtain the multimode field quadrature data. Motivated by standard two-photon SPDC, we started by looking for covariance between each pair of modes. However, we observed no covariance between any pair of the modes in either the two-mode or three-mode case, as shown in Fig.~\ref{PumpSweepCoskewness}. Instead, we found significant coskewness between the signals (see Figs.~\ref{PumpSweepCoskewness}, \ref{InteractionList}).  To verify that the observed coskewness was a coherent process related to the pump, we swept the pump phase and computed the covariances and coskewnesses, as shown in Fig.~\ref{PumpSweepCoskewness} for both the two-mode and three-mode cases. We see that the coskewness oscillates with the pump phase between positive and negative extremes, while the covariance remains unchanged at zero. This observation of coskewness (three-body interactions) in the absence of covariance (two-body interactions) is a striking and extremal observation of non-Gaussian statistics in both the two-mode and three-mode cases. Further, in the three-mode case, it clearly demonstrates three-photon interference, which has only very recently been observed in any system \cite{Hamel2014,Agne2017a}. 

With a total of six quadratures, there are a number of two-mode and three-mode coskewnesses that we can compute. Fig.~\ref{InteractionList} shows several of them.  In the two-mode case, we have a degeneracy between two of the three generated photons, that is, two photons will go to one of the modes. Therefore we expect that only terms such as $\langle I_1^2 I_2\rangle$, where one mode participates twice in the coskewness, will be nonzero. In the three-mode case, where one photon goes to each mode, we expect that the only nonzero coskewness terms should contain all three modes, \textit{e.g.}, $\langle I_1 I_2 I_3 \rangle$. This is exactly what is observed.

In an attempt to impose structure on the myriad three-body correlators, we now look to generalize the phase rotation analysis used in Fig.~\ref{singleModeFig}c. We take as transformations to study the set of symmetry operations of an $N$-mode Gaussian state, which forms the symplectic group \cite{Simon1994} and are generated by quadratic Hamiltonians. The symplectic group includes squeezing operations which create or destroy photons, but we will restrict ourselves to the passive operations of the group.  These operations form a unitary subgroup and include phase rotations of a single mode, with generators such as $\cre{i}\ann{i} + \ann{i}\cre{i}$, and beam-splitter rotations between modes, with generators such as $\cre{i}\ann{j} + \ann{i}\cre{j}$.  Importantly, we expect different cubic Hamiltonians to have different transformation properties under these operations.

With three modes, we have six quadratures, implying a 6D phase space to explore. The symplectic operations can then be represented by $6\times6$ matrices, which can be written as transformations between the quadratures. In order to illustrate the transformation properties with a 3D figure, we project into a three-quadrature subspace. The transformations then become generalized rotations between the three chosen quadratures. Collecting the quadratures into a 3-vector, we can explore arbitrary combinations of the quadratures by apply a series of two rotations according to
\begin{equation}
\left(\ann[A]{}',\ann[B]{}',\ann[C]{}'\right)^t=R_C(\phi)\times R_B(\theta)\times\left(\ann[A]{},\ann[B]{},\ann[C]{}\right)^t,	\label{symOp}
\end{equation}
where $R_i$ are standard 3D rotation matrices with the rotation axis specified by the subscript. After the rotations, we consider the generalized quadrature
\begin{equation}
\ann[A]{\phi\theta}=\cos(\phi)\cos(\theta)\ann[A]{}-\sin(\phi)\ann[B]{}+\cos(\phi)\sin(\theta)\ann[C]{}.	\label{mixQuads}
\end{equation}
and compute its skewness $\gamma_{\phi\theta}$ which in general is a mix of all of the possible three-point correlators of the three quadratures.  Generalizing the one-mode case, we can think of $\gamma_{\phi\theta}$ representing the asymmetry of the 3D distribution of measured quadratures with respect to the symmetry plane perpendicular to the direction of $\ann[A]{ \phi \theta}$.

\begin{figure*}[t]
\center
\includegraphics[width=1\linewidth]{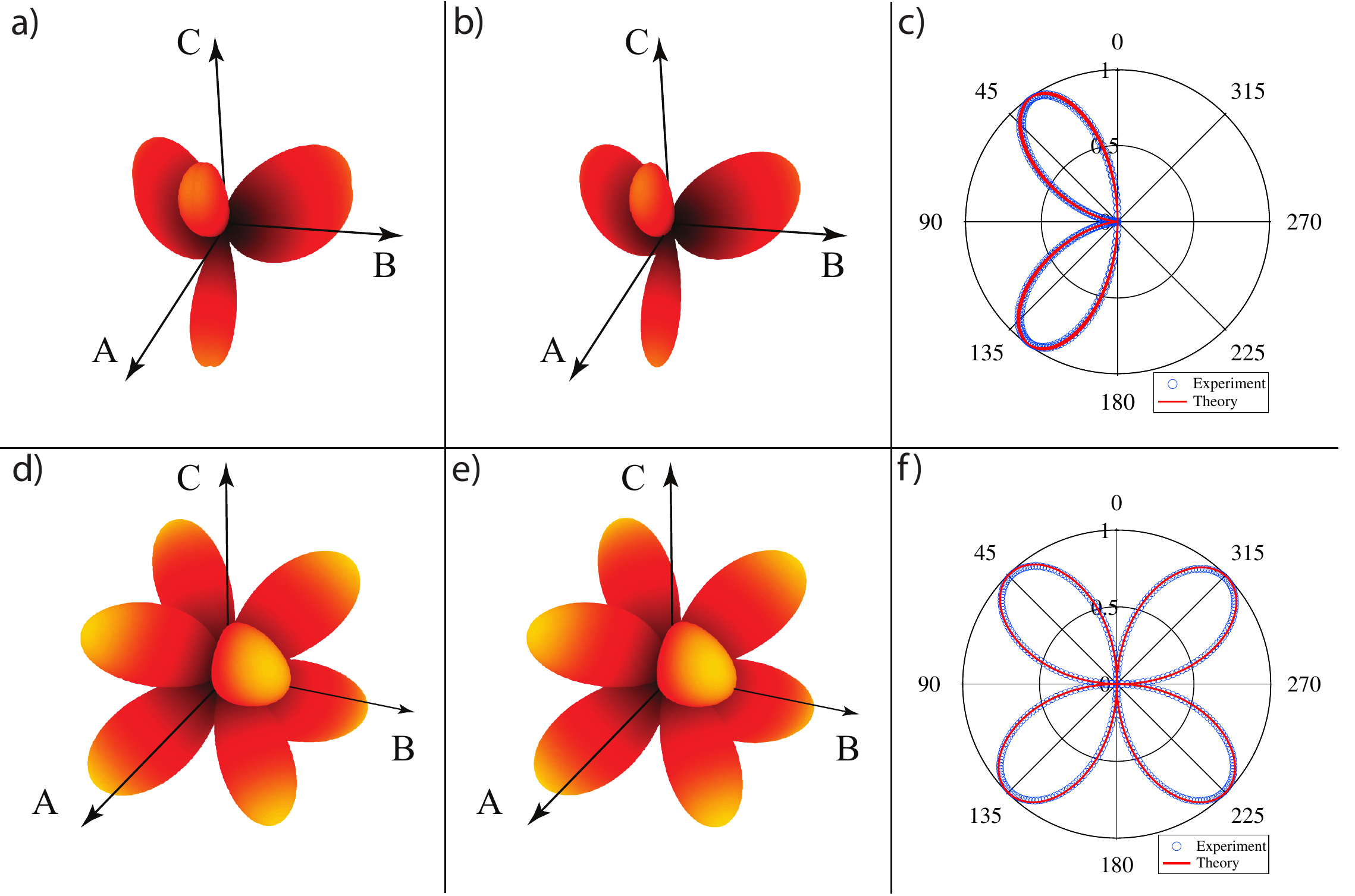}
\caption{Third-order correlation analysis of multimode trisqueezed states. Row 1 and 2 show results for respectively the two-mode and three-mode trisqueezed states, generated by $\hat{H}_{\textrm{2M}}$ and  $\hat{H}_{\textrm{3M}}$. The spherical plots show $\gamma_{\phi\theta}$, which is the skewness of the generalized multimode quadrature $\hat{A}_{\phi\theta}$ (see Eq.~\eqref{mixQuads}), which mixes the mode quadratures through symplectic symmetry operations. The 6D phase space of the three modes is projected into a 3D space for the purposes of visualization.  Generally, $\gamma_{\phi\theta}$ combines the contributions of the skewness and coskewness of the three modes involved in each case. a) Experimental data for the two-mode case. b) The theoretical prediction for the two-mode cubic Hamiltonian $\hat{H}_{\textrm{2M}}$. The clear agreement shows that the observed state is generated by that specific Hamiltonian. c) A plane-cut of the spherical plots through the CA-plane. The curves are normalized, but otherwise the theory has no adjustable parameters. There are no lobes apparent on the right half of the plot because $\gamma_{\phi\theta}$ is negative for $\phi\in[\pi/2,3\pi/2]$, causing these lobes to overlap those of $\phi\in[-\pi/2,\pi/2]$. d) Experimental data for the three-mode case. e) The theoretical prediction for the three-mode cubic Hamiltonian $\hat{H}_{\textrm{3M}}$. f) A plane-cut of the spherical plots through a plane $+35$ degrees from the CA-plane. Again, we see a clear agreement between the observed state and the target Hamiltonian. In particular, the antinodes (lobes) of $\gamma_{\phi\theta}$ appear only at angles where all three modes are mixed, as expected for genuine three-mode interference.
}
\label{multimodeFig}
\end{figure*}

In Fig.~\ref{multimodeFig}, we show spherical plots of $\gamma_{\phi\theta}$ as a function of $\theta\in[0,\pi]$ and $\phi\in[0,2\pi]$ for both the two-mode and three-mode cases. Explicitly, for the two-mode case, we chose $A=I_1$, $B=Q_1$ and $C=I_2$.  The transformation specified by Eq.~\eqref{symOp} is then first a beam-splitter rotation by $\theta$ between mode 1 and mode 2, followed by a phase rotation of mode 1 by $\phi$  (which mixes I and Q). In the three-mode case, we have $A=I_1$, $B=I_2$ and $C=I_3$, such that Eq.\eqref{symOp} describes two beam-splitter rotations among the three modes, first coupling mode 1 and mode 3 and then modes 1 and 2. Fig.~\ref{multimodeFig} shows the experimental results for $\gamma_{\phi\theta}$ along with theoretical predictions for the states produced by the Hamiltonians $\hat{H}_{\textrm{2M}}$ and  $\hat{H}_{\textrm{3M}}$ in Table~\ref{effIntHamTable}.  We remark that the agreement between theory and experiment is very good. We see that different processes produce very different shapes for $\gamma_{\phi\theta}$, which usefully ``fingerprint" the underlying Hamiltonians and allow us to see in a clear, visual way how cleanly we generate one Hamiltonian compared to another.

As an example, we can consider the three-mode case. A first important feature we note is that $\gamma_{\phi\theta}$ has nodes in the AB, BC and CA-planes. We can understand the presence of these nodes by noticing that in these planes at least one of the modes is excluded from the generalized quadrature $\ann[A]{ \phi \theta}$. That is, this tells us that two-mode correlators such as $\langle I_1^2 I_2 \rangle$ and single-mode correlators such as $\langle I_1^3\rangle$ are zero, exactly as we would expect for a state generated by the pure three-mode Hamiltonian $\hat{H}_{\textrm{3M}}$.  Conversely, the antinodes (lobes) correspond to the angles where the contribution of the three-mode correlator $\langle I_1 I_2 I_3 \rangle$ are maximized.  The pattern therefore tells us both that we have activated $\hat{H}_{\textrm{3M}}$ and that we have \textit{not} activated $\hat{H}_{\textrm{1M}}$ nor $\hat{H}_{\textrm{2M}}$.  We note separately that this is strong evidence that we are observing genuine three-mode interference.

We can understand the structure of the two-mode state in Fig.~\ref{multimodeFig}a in a similar way.  First, recall that for our specific choice of pump frequency (see Table~\ref{effIntHamTable}), we expect that we are creating two photons in mode 1 and one photon in mode 2. We can then consider the behavior in the CA-plane, also highlighted in Fig.~\ref{multimodeFig}c. We see clear nodes at $\theta = 0~(\pi/2)$, where we are calculating the single-mode skewness of $\ann[I]{1}$ $(\ann[I]{2})$ alone, in agreement with our expectations. Instead $\gamma_{\phi\theta}$ is maximum around $\theta=\pi/4$ and $\theta=3\pi/4$ where $\ann[I]{1}$ and $\ann[I]{2}$ are maximally mixed, and we get the maximum contribution from the two-mode correlator $\langle I_1^2 I_2 \rangle$. As above, this fingerprint nicely indicates that we are, in this case, activating $\hat{H}_{\textrm{2M}}$ but not $\hat{H}_{\textrm{1M}}$.  We note that the expected lobes at $\theta=5\pi/4$ and $\theta=7\pi/4$ are missing at first glance, but in fact they overlap the lobes at $\theta=\pi/4$ and $\theta=3\pi/4$ because the sign of $\gamma_{\phi\theta}$ becomes negative.  The structure in the AB-plane can be explained in a similar way, except there we are mixing $\ann[Q]{1}$ and $\ann[I]{2}$.

We can  generate several more of these projections of the 6D phase space into 3D, but it is already clear from these two examples that the structure of $\gamma_{\phi\theta}$ is a useful way to characterize the output state.  In particular, with a library of the possible forms generated by different cubic Hamiltonians, we can quickly see which one is generating the observed state.  By observing the relative depth of nodes compared to the antinodes, we can also appreciate how purely we generate just a single member of the cubic Hamiltonian family. For instance, for the three-mode data shown in Fig.~\ref{multimodeFig}d, this ratio is approximately $10^{-4}$, indicating a high degree of purity.

\subsection{Correlation feed-forward}
In this section, we explore more deeply the striking feature that our observed three-photon states exhibit three-mode correlations (skewness) in the absence of two-mode correlations (covariance), which is in strong contrast to the conventional two-photon states and Gaussian states generally. An interesting experimental observation in this direction is that, while the three-mode trisqueezed state does not show covariance between any pairs of modes when starting from a vacuum state, we observed that by seeding one of the three modes with a weak coherent tone the noise power emitted from the other modes is enhanced and that it then has nonzero covariance. Similarly, in the two-mode trisqueezed state, by seeding the mode participating only once, \textit{i.e.} mode 2, we observe the emission of squeezed noise from the other mode. These results can be understood from the point of view of dynamical constraints imposed by the conservation of energy. In the standard two-photon case, signal and idler photon pairs are constrained to have a symmetric detuning around $f_p/2$. In the rotating frame at $f_p/2$, the phasors of the signal and idler photons precess at the same frequency, but in opposite directions, such that the axis (phase) of the sum phasor is stationary in time and the same for all pairs. This gives rise to the observed covariance between the signal and idler modes. In the three-photon case, we have three free frequencies (energies) with no fixed relation between them for any pair of photons, washing out the two-mode correlations. Seeding effectively fixes one photon frequency in the resulting stimulated emission, leaving a fixed relation between the other two. This produces correlations similar to two-photon SPDC.

\begin{figure}[!ht]
\center
\includegraphics[width=0.75\linewidth]{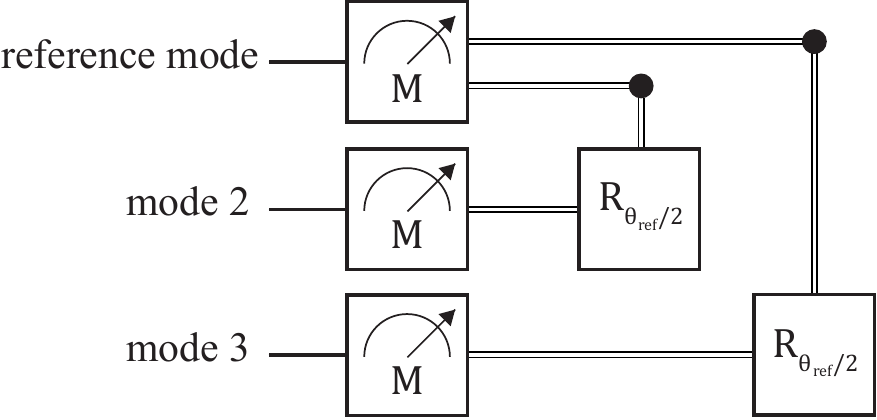}
\caption{A cartoon of the correlation feed-forward protocol for the three-mode trisqueezed state. This scheme tests our hypothesis on the conditional structure of two- and three-mode correlations. In the three-mode protocol shown here, we first measure the local phase of the reference mode (mode 1 in this example), followed by applying a rotation by $\phi_{ref}/2$ on the other two modes. The result is an observable correlation between mode 2 and mode 3 with the characteristic structure of two-photon SPDC, i.e., two-mode squeezing. }
\label{CorrFFDiag}
\end{figure}

\begin{table}[h]
\centering
\begin{tabular}{@{}ccccc@{}}
\toprule
               & \multicolumn{4}{c}{Feed-forward}                                                                                        \\
               & \multicolumn{2}{c}{Yes}                                        & \multicolumn{2}{c}{No}                                         \\ \cline{1-5}
Quads             & \begin{tabular}[c]{@{}c@{}}Pump\\ On\end{tabular} & \begin{tabular}[c]{@{}c@{}}Pump\\ Off\end{tabular} & \begin{tabular}[c]{@{}c@{}}Pump\\ On\end{tabular} & \begin{tabular}[c]{@{}c@{}}Pump\\ Off\end{tabular} \\ \cline{1-5}
\multicolumn{1}{c|}{$I_2I_3$} & $~~0.30\pm0.02$                  & \multicolumn{1}{c|}{$0.00\pm0.02$}          & $0.00\pm0.02$                  & $0.00\pm0.02$                    \\
\multicolumn{1}{c|}{$Q_2Q_3$} & $-0.30\pm0.02$                 & \multicolumn{1}{c|}{$0.00\pm0.02$}          & $0.00\pm0.02$                 & $0.00\pm0.02$                    \\
\multicolumn{1}{c|}{$I_2Q_3$} & $-0.36\pm0.02$                 & \multicolumn{1}{c|}{$0.00\pm0.02$}          & $0.00\pm0.02$                    & $0.00\pm0.02$                    \\
\multicolumn{1}{c|}{$Q_2I_3$} & $-0.36\pm0.02$                 & \multicolumn{1}{c|}{$0.00\pm0.02$}          & $0.00\pm0.02$                    & $0.00\pm0.02$                    \\ \botrule
\end{tabular}
\caption{The resultant two-mode correlation coefficient (normalized covariance) between mode 2 and mode 3 after applying correlation feed-forward to the three-mode trisqueezed state (see Fig.~\ref{CorrFFDiag}). As shown, we recover a significant amount of correlation when applying our feed forward correction. We include the results of applying the protocol to the system noise with the pump off to verify that the protocol itself does not introduce spurious correlations.}
\label{ThreeModeCF}
\end{table}


\begin{figure}[!ht]
\center
\includegraphics[width=0.9\linewidth]{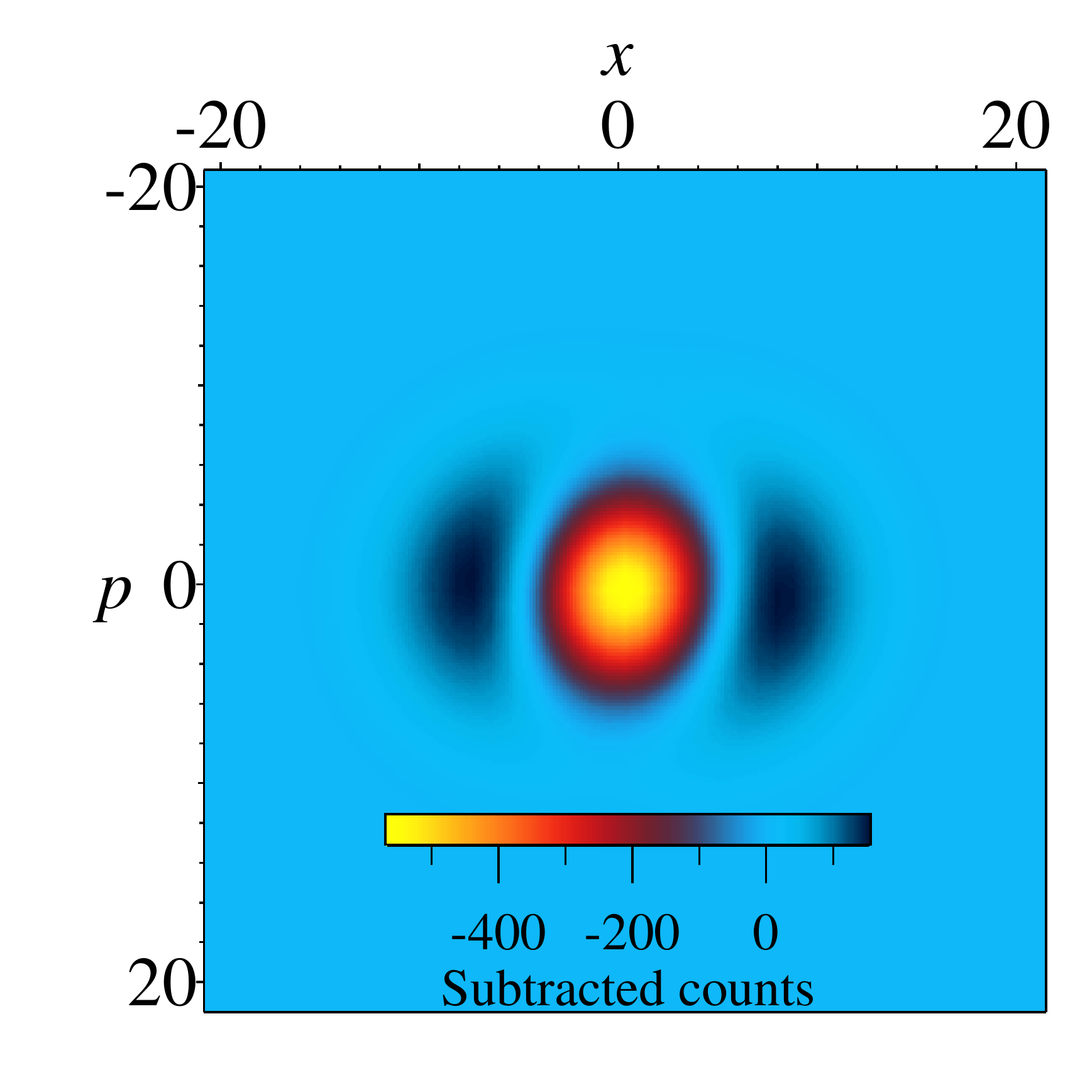}
\caption{The resultant single-mode histogram of mode 1 after applying the correlation feed-forward protocol to the two-mode trisqueezed state. We take mode 2 as the reference mode, correcting the phase of mode 1. Then, we compute the histogram of the data, and subtract from it the histogram of the system noise (with the pump turned off) after also applying the protocol. We clearly see that the subtracted histogram is stretched along the x-axis, indicating squeezing of the ON state. The small rotation of the figure can be explained by a small misalignment of the pump and digitizer phases (see Fig.~\ref{InteractionList}).}
\label{TwoModeCF}
\end{figure}

This argument suggests that by having information about one mode, it should be possible to reveal the two-mode correlation of the remaining modes. That is, by conditioning our measurements of the remaining modes on the measurement of the first ``reference" mode, we should be able recover a conditional distribution with two-mode correlations. To demonstrate this and validate our hypothesis, we perform the following ``correlation feed-forward" protocol. First, we estimate the phase of the reference mode for every sample period using the standard relation $\phi_{ref}=\tan^{-1}(Q_{ref}/I_{ref})$.  (Our data was sampled at 1 MHz, corresponding to an integration time of 1 $\mu$s.) We then rotate the quadratures of the remaining modes using $\phi_{ref}$, as illustrated in Fig.~\ref{CorrFFDiag}. The action of the phase rotation can again be explained by energy conservation: as the frequencies of the three photons must sum to $f_p$, a small frequency shift $(\delta f_{ref})$ from the center in the reference mode needs to be compensated by changes in the other two modes.  In the three-mode case, we used mode 1 as the reference and applied rotations by $\phi_{1}/2$ to modes 2 and 3.  Table~\ref{ThreeModeCF} shows the resulting correlations recovered. 

We also applied the protocol to the two-mode trisqueezed state, using mode 2 as reference. After doing this, we see squeezing effects in mode 1 with a ratio in the variance of $\ann[x]{}$ to $\ann[p]{}$ of $1.11$, compared to $1.00$ without the feed-forward.  To visualize the effect of the correlation feed-forward protocol, in Fig.~\ref{TwoModeCF} we show a histogram of the corrected quadratures after subtracting the system noise histogram. We can clearly see a stretching of the distribution along the $\ann[x]{}$ quadrature, indicating squeezing-like correlations.  

These results validate our hypothesis about the conditional structure of two-mode and three-mode correlations in our system. At the same time, the correlation feed-forward demonstrations here are only proofs of principle because our phase measurements are strongly contaminated by the system noise.  In future work, it would be interesting to explore the fundamental limits of this reconstruction method.
  


\section{Conclusion}
In this work, we demonstrated a device which implements efficient three-photon SPDC in the microwave domain. The device is highly flexible, allowing us to perform single-mode, two-mode and three-mode SPDC by simply selecting the appropriate pump frequency. We carefully compared the observations of various features of the three-photon SPDC states, verifying that our output signal was generated by the chosen interaction Hamiltonian. By extending parametric interactions to higher order, our device has opened up novel possibilities that will enable a new wave of novel experimental and theoretical studies in microwave quantum optics.

\section*{ACKNOWLEDGEMENT}
The authors wish to thank B. Plourde, J.J. Nelson and M. Hutchings at Syracuse University for invaluable help in junction fabrication. We also thank J. Aumentado of NIST-Boulder for providing the SNTJ calibration source. Finally, we thank B. Sanders for useful discussions and insight. CMW, CWSC, IN and AMV acknowledge the Canada First Research Excellence Fund (CFREF), NSERC of Canada, the Canadian Foundation for Innovation, the Ontario Ministry of Research and Innovation, and Industry Canada for financial support. CS has received financial support through the Postdoctoral Junior Leader Fellowship Programme from Òla CaixaÓ Banking Foundation (LCF/BQ/LR18/11640005).  P. F.-D. acknowledges support of a fellowship from ``la Caixa Foundation" (ID100010434) with code LCF/BQ/PR19/11700009, and funding from the Ministry of Economy and Competitiveness, through contracts FIS2017-89860-P and Severo Ochoa SEV-2016-0588. FQ and GJ acknowledge the Knut and Alice Wallenberg Foundation.

\end{document}